\documentclass{jpsj-suppl}
\usepackage{times}
\usepackage{color}
\usepackage{ulem}

\title{
Monte Carlo Study of an Effective Ising Model for\\
the Spin-ice type Kondo Lattice Model
}

\author{
Hiroaki Ishizuka\thanks{E-mail address: ishizuka@aion.t.u-tokyo.ac.jp}, Masafumi Udagawa, and Yukitoshi Motome
}
\inst{
Dept. of Applied Physics, the University of Tokyo, 7-3-1 Hongo, Bunkyo, Tokyo 113-8656
}
\abst{
An effective Ising model for the spin-ice type Kondo lattice model is investigated by the classical Monte Carlo simulation.
We clarify the magnetic phase diagram with four phases: ice-ferro, ice-$(0,0,2\pi)$, 32-sublattice, and all-in/all-out ordered states.
The result well reproduces the phase diagram of the Kondo lattice model studied previously [H. Ishizuka {\it et al}.: J. Phys. Soc. Jpn. {\bf 81} (2012) 113706], which suggests that the RKKY interactions up to third neighbors are sufficient to describe the magnetic properties of the itinerant electron model.
We discuss the peculiar nature of phase transitions: the suppression of the critical temperatures down to zero temperature between two ice phases and the presence of the tricritical points on the phase boundary of the 32-sublattice ordered state.
}

\kword{
geometrical frustration, pyrochlore lattice, spin ice, RKKY interaction, Monte Carlo simulation
}

\begin{document}
\maketitle

\section{Introduction}

Frustration in magnetic materials offers a fertile ground for studying interesting phenomena in strongly correlated systems~\cite{Diep,Lacroix}. 
Competing interactions under frustration often lead to an extensive number of energetically degenerate states.
Even a small perturbation to the degeneracy can result in remarkable effects, such as phase transitions and colossal responses to external fields.
These unique properties have stimulated intensive studies of competing orders and fluctuations in frustrated systems. 

\begin{figure}
\begin{center}
\includegraphics[width=\linewidth]{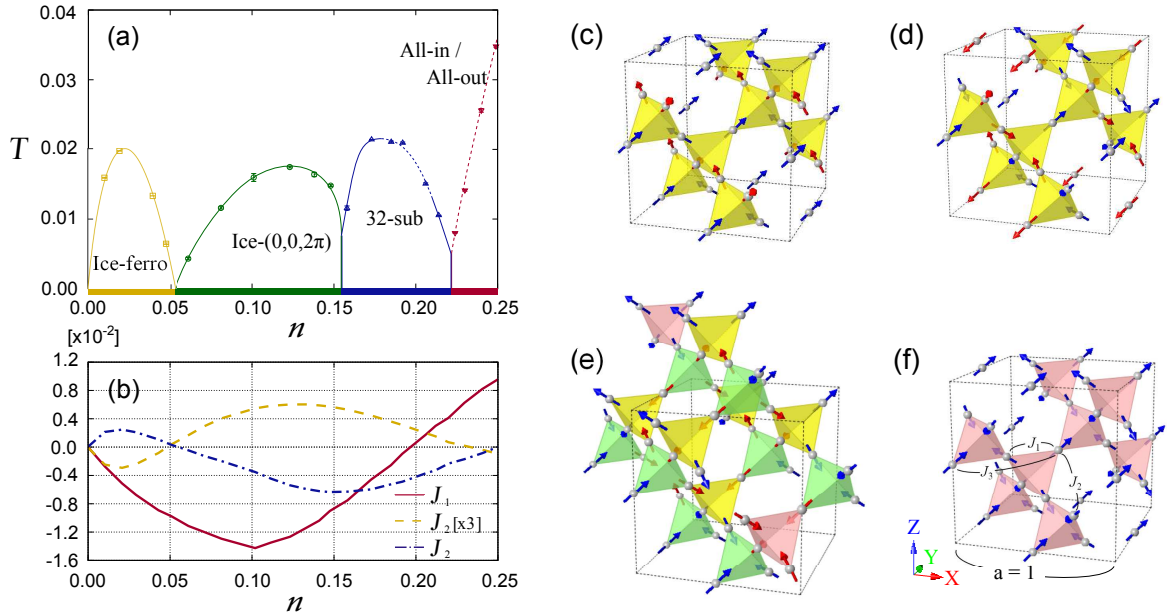}
\end{center}
\caption{(Color online)
(a) Phase diagram of the effective RKKY Ising model in Eq.~(\ref{eq:Hspin}).
The lines are the guides for eyes; the solid (dotted) lines show the first(second)-order transitions.
The bottom strip shows the ground state phase diagram obtained by a variational calculation comparing the ground state energy between the different magnetic states.
(b) The RKKY interactions for nearest-, second-, and third-neighbor pairs.
Figures (c)-(f) show schematic pictures of the magnetic orders in the four phases present in the phase diagram: (c) ice-ferro, (d) ice-($0,0,2\pi$), (e) 32-sublattice, and (f) all-in/all-out orders.
}
\label{fig:diagram}
\end{figure}

One such example of frustrated systems is spin ice~\cite{Harris1997,Ramirez1999,GingrasPreprint}.
In spin ice, spins with strong Ising-type anisotropy along the sublattice-dependent local $[111]$ direction reside on the pyrochlore lattice, which consists of corner-sharing tetrahedra [see Figs.~\ref{fig:diagram}(c)-\ref{fig:diagram}(f)].  
In the spin-ice compounds, the effective interaction for nearest-neighbor (NN) spins becomes ferromagnetic, which enforces a two-in two-out spin configuration in each tetrahedron~\cite{Siddharthan1999,Bramwell2001,denHertog2005}.
The local constraint, called the ice rule~\cite{Pauling1935}, is not sufficient to establish a long-range order, and results in the
macroscopic degeneracy in the ground state.
The long-range dipolar interactions lead to significant changes in the phase diagram~\cite{Melko2004} and emergence of peculiar excitation called magnetic monopoles~\cite{Castelnovo2008}. 

On the other hand, recently, the physics of spin-charge coupling in geometrically frustrated systems has also gained interest.
These studies were stimulated by recent experiments on the Mo and Ir metallic pyrochlore oxides which show rich phase diagrams~\cite{Hanasaki2007,Iguchi2009,Matsuhira2011}.
Indeed, some of their transport properties were theoretically discussed in the context of spin-charge coupling~\cite{Motome2010a,Motome2010b,Udagawa2012}.
At the same time, the magnetism of a pyrochlore Ising model with the Ruderman-Kittel-Kasuya-Yosida (RKKY)~\cite{Ruderman1954,Kasuya1956,Yosida1957} interactions~\cite{Ikeda2008} and a Kondo lattice model on a pyrochlore lattice~\cite{Ishizuka2012} were studied.
In the later study, the magnetic phase diagram of the Kondo lattice model was mapped out by using a Monte Carlo (MC) simulation, and it was pointed out that the change of the RKKY interactions depending on the electron density plays an important role in understanding the phase diagram.
However, the detailed investigation on the effective RKKY Ising model was not presented.

In this contribution, we numerically study the magnetic phase diagram of the effective Ising model with long-range RKKY interactions using a MC simulation.
We present the phase diagram of the effective model and show that the result well reproduces the phase diagram of the Kondo lattice model studied in Ref.~\citen{Ishizuka2012}.
In addition, we investigate the critical behavior for the phase transition to each magnetic phase, which was unclear in the previous study on the Kondo lattice model due to the computational limitation.

\section{Model and Method}
\subsection{Model}
The Hamiltonian for the spin-ice type Kondo lattice model considered in the previous studies~\cite{Udagawa2012,Ishizuka2012} is given by
\begin{eqnarray}
H = -t \! \sum_{\langle i,j \rangle, \sigma} \! ( c^\dagger_{i\sigma} c_{j\sigma} + \text{H.c.} ) -J \sum_{i} {\bf S}_i \cdot {\boldsymbol \sigma_i}. \label{eq:Hkondo}
\end{eqnarray}
Here, the first term is the hopping of itinerant electrons, where $c_{i\sigma}$ ($c^\dagger_{i\sigma}$) is the annihilation (creation) operator of an itinerant electron with spin $\sigma= \uparrow, \downarrow$ at $i$th site.
The sum $\langle i,j \rangle$ is taken over NN sites on the pyrochlore lattice, and $t$ is the NN transfer integral.
The second term is the onsite interaction between localized spins and itinerant electrons, where ${\bf S}_i$ and ${\boldsymbol \sigma}_i$ represent the localized Ising spin and itinerant electron spin at $i$th site, respectively ($|{\bf S}_i|=1$), and $J$ is the coupling constant (the sign of $J$ does not matter in the present model as the sign of $J$ can be reversed by time-reversal transformation).
The anisotropy axis of the Ising spin is given along the local [111] direction at each site, i.e., along the line connecting the centers of two tetrahedra to which the spin belongs [see Figs.~\ref{fig:diagram}(c)-\ref{fig:diagram}(f)].

In the model in Eq.~(\ref{eq:Hkondo}), the kinetic motion of electrons induces effective magnetic interactions between the localized Ising spins. 
In the weak coupling limit of $J/t \ll 1$, the effective interactions are obtained by the second-order perturbation theory in terms of the second term in Eq.~(\ref{eq:Hkondo}).
They are called the RKKY interactions~\cite{Ruderman1954,Kasuya1956,Yosida1957}. 
Thus, the perturbation gives an effective Ising spin model with long-range RKKY interactions. 
For simplicity, by omitting the interactions further than third neighbors and projecting the local spin axis along the direction of the anisotropy axis~\cite{Moessner1998}, we consider the Hamiltonian in the form
\begin{eqnarray}
H = - J_1 \sum_{\langle i,j \rangle} S_i^z S_j^z - J_2 \sum_{\left\{ i,j \right\}} S_i^z S_j^z - J_3 \sum_{\left[ i,j \right]} S_i^z S_j^z, \label{eq:Hspin}
\end{eqnarray}
where $S_i^z=\pm1$ is the projected collinear Ising moment at $i$th site and the sum $\left\{ i,j \right\}$ ($\left[ i,j \right]$) is taken over second(third)-neighbor pairs.
Here, $J_1$, $J_2$, and $J_3$ are the nearest-, second-, and third-neighbor interactions, respectively, which are dependent on the electron density $n=\frac1{2N}\sum_{i\sigma} \langle c_{i\sigma}^\dagger c_{i\sigma}\rangle$ in the original Kondo lattice model in Eq.~(\ref{eq:Hkondo}); 
the estimates are shown in Fig.~\ref{fig:diagram}(b) as functions of $n$ (we set $J^2/t=1$ as the energy unit hereafter).
We call this model the effective RKKY Ising model.
Note that the previous study on the similar model assumed the RKKY interaction for free electrons~\cite{Ikeda2008}, while our effective model takes account of the band structure of the pyrochlore lattice.
In the following, we take the lattice constant of cubic unit cell $a = 1$ [see Fig.~\ref{fig:diagram}(f)], and the Boltzmann constant $k_{\rm B} = 1$.

\subsection{Monte Carlo method}

We investigate the phase diagram of the effective RKKY Ising model in Eq.~(\ref{eq:Hspin}) by a classical MC simulation for $J_1$, $J_2$, and $J_3$ at each $n$ plotted in Fig.~\ref{fig:diagram}(b).
The single-spin flip update by using the heat-bath method was employed for the MC sampling. 
Most of the calculations were initially started from random spin configurations,
while the calculations in the ice-(0,0,2$\pi$) region were started from a mixed initial spin-configuration of low-temperature ($T$) ordered and high-$T$ disordered states in order to deal with the severe freezing at low $T$.
The typical system sizes for the calculations were $N=4\times6^3$ to $4\times12^3$ sites, whereas the calculations up to $N=4\times24^3$ were done for the cases that weak first order transitions were expected.
The calculations were typically done with $5\times10^6$ MC steps after the thermalization of $10^6$ MC steps, and the error bars were estimated by dividing the MC data into five bins and calculating the corrected sample standard deviation among the bins.

\section{Results}

\subsection{Phase diagram}

Figure~\ref{fig:diagram}(a) shows the phase diagram for the effective RKKY Ising model in Eq.~(\ref{eq:Hspin}) obtained by MC calculations.
We identify four magnetic phases at low $T$: (i) the ice-ferro state for $n\lesssim 0.054$, (ii) the ice-$(0,0,2\pi)$ state for $0.054\lesssim n \lesssim 0.150$, (iii) the 32-sublattice ordered state for $0.150\lesssim n \lesssim 0.222$, and (iv) all-in/all-out ordered state for $0.222 \lesssim n$.
The schematic pictures of each magnetic order are shown in Figs.~\ref{fig:diagram}(c)-\ref{fig:diagram}(f).
The symbols in the phase diagram show the critical temperatures $T_c$ estimated from $T$ dependence of the order parameters (see the next section for the details).
The strip at the bottom of the figure is the ground state phase diagram obtained by a variational calculation comparing the ground state energy for the four magnetic orders.
The variational estimates of the range of $n$ for each phase are in good agreement with the MC results, as shown in Fig.~\ref{fig:diagram}(a). 

The magnetic phase diagram in Fig.~\ref{fig:diagram}(a) is in good accordance with the phase diagram for the Kondo lattice model in Eq.~(\ref{eq:Hkondo}) obtained in Ref.~\citen{Ishizuka2012}.
Indeed, all the four magnetic states appeared in the Kondo lattice model at $J/t=2$ are found in the effective RKKY Ising model.
The range of $n$ for each phase is in good agreement between the two models, except for the electronic phase separation specific to the Kondo lattice model.
The relative values of $T_c$ for different phases also show reasonable accordance between the two models, while the magnitude of $T_c$ is overestimated by a factor of 2-3 in the present results when considering the value of $J/t=2$ taken in the Kondo lattice model.
The results indicate that the effective model with the RKKY interactions up to third neighbors semi-quantitatively describes the magnetic properties of the original Kondo lattice model in the weak coupling regime.

Although it was unclear how the ice-ferro and ice-$(0,0,2\pi)$ ordered phases meet with each other at low $T$ in the previous study for the Kondo lattice model~\cite{Ishizuka2012}, our result in Fig.~\ref{fig:diagram}(a) suggests that the two phase boundaries go to zero and meet at $T=0$.
This peculiar behavior can be understood as follows.
The ground state energies per site for the ice-ferro and ice-$(0,0,2\pi)$ ordered states are given by $E_\text{if}=4J_1+8J_2-12J_3$ and $E_{\rm 2\pi}=4J_1-8J_2+4J_3$, respectively.
At the boundary between the two states, $E_\text{if} = E_{\rm 2\pi}$; this implies $J_2=J_3$ at the phase boundary.
This is indeed the case, as shown in Fig.~\ref{fig:diagram}(b). 
It was recently pointed out that the model in Eq.~(\ref{eq:Hspin}) with $J_2=J_3$ can be rewritten by introducing the magnetic charge for each tetrahedron $p$, $Q_p=\sum_{i\in p}S_i^z$, into the form of~\cite{Ishizuka2013}
\begin{eqnarray}
H=-J_2 \sum_{\langle p, q\rangle} Q_p Q_q - \left( \frac{J_1}4 - \frac{J_2}2 \right) \sum_p Q_p^2 + {\rm const}.
\end{eqnarray}
Here, $p$ and $q$ are the indices for the tetrahedra in the pyrochlore lattice;
the first sum is taken over all the pairs of NN tetrahedra, and the second sum over all tetrahedra.
Rewriting the Hamiltonian in this form, it is easily seen that all the two-up two-down spin configurations are energetically degenerate since $Q_p=0$ for all $p$.
Hence, when $J_1$ is dominantly negative and favors the two-up two-down configurations, the system remains disordered to $T\to0$ when $J_2=J_3$.
This explains the reason why $T_c$ goes to zero at the phase boundary between the ice-ferro and ice-($0,0,2\pi)$ phases. 

\subsection{Temperature dependence of physical quantities}

\begin{figure}
\begin{center}
\includegraphics[width=0.86\linewidth]{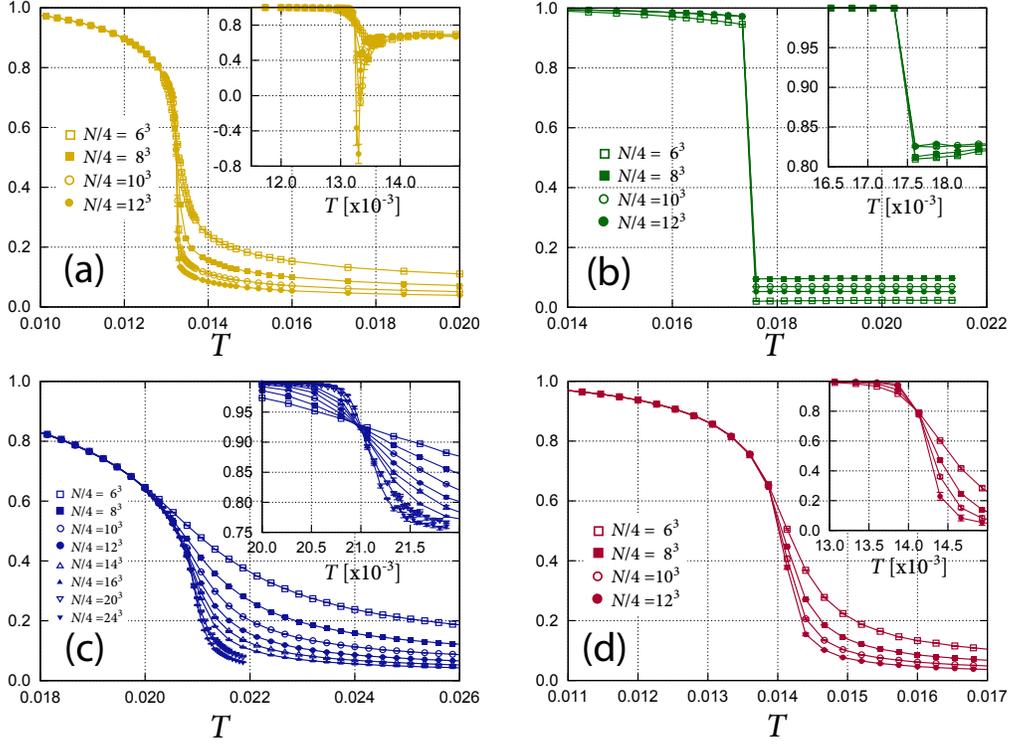}
\end{center}
\caption{(Color online)
Temperature dependence of the order parameters in each phase: (a) $m_\text{if}$ at $n=0.039$, (b) $m_{\rm 2\pi}$ at $n=0.123$, (c) $m_\text{32}$ at $n=0.192$, and (d) $m$ at $n=0.230$.
The insets in each figure show the temperature dependence of the Binder parameters for each parameter in the main panel.
}
\label{fig:mc}
\end{figure}

Let us next look into the critical behavior of the phase transition for each phase while changing $T$.
The temperature dependence of the order parameters for each magnetic phase is shown in Fig.~\ref{fig:mc}.

Figure~\ref{fig:mc}(a) shows the result of the order parameter for the ice-ferro state, $m_\text{if}=\langle \{ 3\sum_i S_{ii}({\bf 0}) - 2\sum_{i>j}S_{ij}({\bf 0}) \}/4N \rangle$, at $n=0.039$; the inset shows the result of the Binder parameter~\cite{Binder1981} for $m_\text{if}$.
Here, $S_{ij}({\bf q})$ ($i,j=0,1,2,3$) is the spin structure factor for the Ising spins on $i$th and $j$th sublattices.
The rapid increase of $m_\text{if}$ around $T_c=0.013$ indicates the phase transition to the ice-ferro state.
At the same time, the Binder parameter for large system sizes $N\ge 4\times 10^3$ shows a dip and takes a negative value near $T_c$, indicating that the transition is of first order.
$T_c$ plotted in Fig.~\ref{fig:diagram}(a) is estimated from the position of the dip for the Binder parameter (the finite size effect is negligibly small).
The discontinuous transition is consistent with the result in the previous study using RKKY interactions for free electron gas~\cite{Ikeda2008}.
In the previous study, the reason for the discontinuity was discussed in relation to the six-state models.
On the other hand, our previous result on the Kondo lattice model appeared to show a continuous change of the order parameter at the phase transition.
This is presumably due to the small system size. 
Indeed, the dip of the Binder parameter does not appear for small sizes in Fig.~\ref{fig:mc}(a).

With increasing the electron density, a sharp first order transition to ice-$(0,0,2\pi)$ ordered state is seen.
Figure~\ref{fig:mc}(b) and its inset show the result of $m_{\rm 2\pi}^2= \frac1{4N} [ 2\sum_{i,j} S_{ii}({\bf q}_j^{2\pi}) -4 \{ S_{01}({\bf q}_1^{2\pi}) + S_{23}({\bf q}_1^{2\pi})  + S_{02}({\bf q}_2^{2\pi})+ S_{13}({\bf q}_2^{2\pi}) + S_{03}({\bf q}_3^{2\pi}) + S_{12}({\bf q}_3^{2\pi}) \} ]$ and its Binder parameter at $n=0.123$, respectively.
Here, ${\bf q}_1^{2\pi}=(0,\pi,\pi)$, ${\bf q}_2^{2\pi}=(\pi,0,\pi)$, and ${\bf q}_3^{2\pi}=(\pi,\pi,0)$.
The abrupt jump of $m_{\rm 2\pi}$ at $T_c=0.0524(4)$ clearly indicates that the transition is of first order.
As the system size dependence of $T_c$ is negligible, we used $T_c$ for $N=4\times 12^3$ in the plot in Fig.~\ref{fig:diagram}(a).
The discontinuous transition is consistent with the result in the Kondo lattice model in Ref.~\citen{Ishizuka2012}.

On the other hand, the phase transition to the all-in/all-out state in the higher density region is second order. 
Figure~\ref{fig:mc}(d) shows $T$ dependence of $m=\sum_{i,j}S({\bf q}={\bf 0})/4N$ at $n=0.230$. 
$m$ shows continuous increase with decreasing $T$, and the Binder parameter for different sizes shows a crossing at a temperature [the inset of Fig.~\ref{fig:mc}(d)], indicating the second order transition; $T_c$ is determined from the crossing of the Binder parameter shown in the inset.
The continuous phase transition was also seen for the Kondo lattice model in Ref.~\citen{Ishizuka2012}.

A second order transition also takes place in the phase transition to the 32-sublattice ordered state.
Figure~\ref{fig:mc}(c) shows $T$ dependence of $m_{32}=\langle \sum_i S_{ii}({\bf q}_i^{32})/N \rangle$ at $n=0.192$, where ${\bf q}_0^{32}=(\pi,\pi,\pi)$, ${\bf q}_1^{32}=(\pi,0,0)$, ${\bf q}_2^{32}=(0,\pi,0)$, and ${\bf q}_3^{32}=(0,0,\pi)$.
The results of the Binder parameters for $m_{32}$ are also shown in the inset.
Smooth increase of $m_{32}$ and the crossing of the Binder parameter for different system sizes indicate a second order transition.
Up to the calculation of system sizes $N=4\times 24^3$, the second order transition takes place for electron densities $0.18\lesssim n \lesssim 0.21$ as shown by the dotted lines in Fig.~\ref{fig:diagram}(a).
On the other hand, by the Binder analysis, the phase transition to the 32-sublattice ordered state becomes first order for $0.15\lesssim n \lesssim 0.18$ and $0.21\lesssim n \lesssim 0.22$.
This implies the presence of the tricritical points on the phase boundary for the 32-sublattice ordered phase~\cite{note_32sub}.

On the other hand, in the study on the Kondo lattice model, no first order transition was observed~\cite{Ishizuka2012}.
The difference is presumably due to the finite size effect.
Indeed, in the current calculations, we found that the system size above $N=4\times 12^3$ is necessary for distinguishing the first order transition.
On the other hand, due to the computational time, previous calculations on the Kondo lattice model were done with $N=4\times 8^3$ at the largest.

\section{Summary}

To summarize, we numerically investigated an effective RKKY Ising model for the spin-ice Kondo lattice model, which was studied in Ref.~\citen{Ishizuka2012}, by using the Monte Carlo simulation.
We showed that the effective model with up to the third-neighbor RKKY interaction well reproduces the phase diagram of the Kondo lattice model.
In addition, we presented the detailed results on the nature of the phase transition to each magnetically ordered state.
We showed that the critical temperature goes to zero between the two ice phases, and that the tricritical points are present on the phase boundary of the 32-sublattice ordered state.

\section*{Acknowledgement}

H.I. is supported by Grant-in-Aid for JSPS Fellows.
This research was supported by KAKENHI (No. 21340090, 22540372, 24340076, and 24740221), the Strategic Programs for Innovative Research (SPIRE), MEXT, and the Computational Materials Science Initiative (CMSI), Japan.


\begin{thebibliography}{99}
\bibitem{note_32sub}      In the limit of $J_1\to0$, $J_2\to0$, and $J_3<0$, the transition to 32-sublattice ordered state is essentially the same with a Neel transition on a cubic lattice.
Hence, the tricritical point is expected in the $J_1$-$J_2$-$J_3$ Ising model on a pyrochlore lattice.
\end{thebibliography}
\end{document}